\documentclass[%
reprint,
superscriptaddress,
showpacs,
 amsmath,amssymb,
 aps,
]{revtex4-2}

\usepackage[utf8]{inputenc}

\usepackage{graphicx}
\usepackage{dcolumn}
\usepackage{bm}
\usepackage{mathrsfs} 
\usepackage{physics}
\usepackage{booktabs}
\usepackage{subcaption}
\usepackage[dvipsnames]{xcolor}

\usepackage{float}

\usepackage[]{hyperref}
\usepackage{tikz}
\usetikzlibrary{arrows,shapes}
\usetikzlibrary{trees}
\usetikzlibrary{matrix,arrows} 				
\usetikzlibrary{positioning}				
\usetikzlibrary{calc,through}				
\usetikzlibrary{decorations.pathreplacing}  
\usepackage{pgffor}							
\usetikzlibrary{decorations.pathmorphing}	
\usetikzlibrary{decorations.markings}
\tikzset{
	vector/.style={decorate, decoration={snake}, draw},
	provector/.style={decorate, decoration={snake,amplitude=2.5pt}, draw},
	antivector/.style={decorate, decoration={snake,amplitude=-2.5pt}, draw},
	fermion/.style={draw=black, postaction={decorate},
	decoration={markings,mark=at position .55 with {\arrow[draw=black]{>}}}},
	fermionbar/.style={draw=black, postaction={decorate},
	decoration={markings,mark=at position .55 with {\arrow[draw=black]{<}}}},
	fermionnoarrow/.style={draw=black},
	gluon/.style={decorate, draw=black,
	decoration={coil,amplitude=4pt, segment length=5pt}},
	scalar/.style={dashed,draw=black, postaction={decorate},
	decoration={markings,mark=at position .55 with {\arrow[draw=black]{>}}}},
	scalarbar/.style={dashed,draw=black, postaction={decorate},
	decoration={markings,mark=at position .55 with {\arrow[draw=black]{<}}}},
	scalarnoarrow/.style={dashed,draw=black},
	electron/.style={draw=black, postaction={decorate},
	decoration={markings,mark=at position .55 with {\arrow[draw=black]{>}}}},
	bigvector/.style={decorate, decoration={snake,amplitude=4pt}, draw},
}

\def\Cd{ \mathcal{I}_\mathrm{d} }
\def\Cm{ \mathcal{I}_\mathrm{m} }
\def\Cod{ \mathcal{I}_\mathrm{od} }

\begin{document}


\title{
	$B^0-\bar B^0$ entanglement for an ideal experiment
	on the direct CP violation $\phi_3/\gamma$ phase
}

\author{Jos\'e Bernab\'eu}
\author{Francisco J. Botella}
\author{Miguel Nebot}
\affiliation{%
	Department of Theoretical Physics, University of Valencia
	and IFIC, Univ.~Valencia - CSIC, Burjassot, Valencia, Spain
}%
\author{Alejandro Segarra}%
\affiliation{%
	Institut für Theoretische Teilchenphysik (TTP), Karlsruher Institut für Technologie (KIT), 
	76131 Karlsruhe, Germany
}%

\date{\today}

\begin{abstract}
$B^0$--$\bar B^0$ Entanglement offers a conceptual alternative to the single charged B-decay asymmetry for the measurement of the direct CP violating $\gamma/\phi_3$ phase. With $f =J/\Psi K_L,J/\Psi K_S$ and $g = (\pi\pi)^0,(\rho_L\rho_L)^0$ the 16
time-ordered double decay rate Intensities to $(f, g)$ depend
on the relative phase between the the $f$- and $g$-decay amplitudes 
given by $\gamma$ at tree-level. 
Several constraining consistencies appear. An intrinsic accuracy of the method at the level of
$\pm 1^\circ$ could be achievable at Belle-II with an improved  
determination of the penguin amplitude to $g$-channels from
existing facilities.
\end{abstract}

\maketitle


There is considerable interest in improving the precision 
for the direct CP violation $\phi_3/\gamma$ phase
\mbox{$\gamma = \arg(-V_{ud}^{} V_{ub}^*/V_{cd}^{} V_{cb}^*)$} in the $b$-$d$ unitarity triangle
of the Cabibbo-Kobayashi-Maskawa (CKM) flavour mixing matrix of quarks~\cite{Cabibbo:1963yz, Kobayashi:1973fv}. 
This angle connects the sides for decay amplitudes of the $B$ system dominated by tree diagrams, 
so that its measurement is a bona-fide determination of the Standard Model (SM) parameters. 
This is important in order to search for New Physics in the loop contributions of penguin ---rare decays--- 
and box ---mixing--- diagrams. 

The most precise result from a single analysis at the LHCb experiment is~\cite{LHCb:2021dcr}
$\gamma = (65.4^{+3.8}_{-2.2})^\circ$. 
It uses the GGSZ method~\cite{Giri:2003ty} for \mbox{$B^\pm \to D K^\pm$}
with the choice of \mbox{$D \to K_S \pi^+ \pi^-$}, \mbox{$D \to K_S K^+ K^-$} 3-body decays. 
In charged $B$ decays, 
the observation of CP violation (CPV) needs~\cite{Bernabeu:1980ke} the interference 
of two amplitudes with different weak phases ---changing sign from particles to antiparticles--- 
and strong phases ---invariant under CP---. 
This mismatch is what originates a CP-violating asymmetry in the corresponding decay rates for $B^+$ and $B^-$.
In this case $D$ represents a $D^0$ or $\bar D^0$ meson reconstructed from a final state that is common to both,
$D^0$ and $\bar D^0$ being produced respectively 
by $b \to c \bar u s$ and $b \to u \bar c s$ tree level diagrams. The parameters of their mixing have been simultaneously determined in the analysis of ~\cite{LHCb:2021dcr}.
The amplitude of the decay $B^- \to D K^-$, $D \to K_S h^+ h^-$ can be written as 
a sum of $B^- \to D^0 K^-$ and $B^- \to \bar D^0 K^-$ contributions as
\begin{equation}
	A_B(m^2_-,m^2_+) = A_D(m^2_-,m^2_+) + r_B\, e^{i (\delta_{B}-\gamma)}\, \bar A_D (m^2_-,m^2_+) \,,
\end{equation}

\noindent
where $m^2_\pm$ are the squared invariant masses of the $K_S h^\pm$ particle combinations, 
that define the position of the decay in the Dalitz plot. 
The parameter $r_B^{}$ is the ratio of the magnitudes of the 
\mbox{$B^- \to \bar D^0 K^-$} and \mbox{$B^- \to D^0 K^-$} amplitudes, 
$\delta_B$ being their strong relative phase. 
Neglecting CP violation in charm decays, 
the charge-conjugated amplitudes for $B^+$ decay satisfy $\bar A_D = A_D$. 
The sensitivity to $\gamma$ is obtained by comparing the distributions in the Dalitz plots 
of $D$ decays from $B^+$ and $B^-$ mesons.
As a consequence, the variation of the strong phase within the Dalitz plot is needed. 
Complementary information from measurements performed by CLEO~\cite{Libby:2010nu}
and BESIII~\cite{Ablikim:2020yif, Ablikim:2020lpk, Ablikim:2020cfp} is available.
Alternative methods~\cite{Gronau:1990ra, *Gronau:1991dp, Atwood:2000ck, Grossman:2002aq} 
correspond to different choices for the decay channels of the $D$'s.
One understands the complexity of the analyses.

In this paper we discuss an ideal conceptual experiment for $\gamma$
by exploiting the $B^0-\bar B^0$ Einstein-Podolsky-Rosen (EPR) Entanglement~\cite{Einstein:1935rr}. 
The use of the EPR correlation was proposed in 
Refs.~\cite{Wolfenstein:1983cx, Gavela:1985dw, Falk:2000ga, Alvarez:2003kh} 
for several decay channels in the $B$ factories. 
Entanglement has been instrumental in the past for 
the observation of Time-Reversal-Violation by the BABAR Collaboration~\cite{Lees:2012uka} 
using the concept and method given in Refs.~\cite{Banuls:1999aj, Banuls:2000ki, Bernabeu:2012ab}
and with far reaching information~\cite{Applebaum:2013wxa, Bernabeu:2013qea, Bernabeu:2016sgz}.
The method for $\gamma$ consists in the observation of 
the coherent double decay to flavor-non-specific products.
In it, the extraction of the $\gamma$ phase is free from the essential strong phases contamination 
needed in charged B decays.
The necessary interference between amplitudes containing the $V_{cd}^{} V_{cb}^*$ and $V_{ud}^{} V_{ub}^*$
sides of the unitarity triangle is automatic from the two terms of the entangled 
$B^0-\bar B^0$ system. 
The double rate intensity to the $(f, g)$ and $(g, f)$ pairs of CP-eigenstate decay products,
with $f = J/\psi K_S,\, J/\psi K_L$ and
$g = \pi^+ \pi^-,\, \pi^0 \pi^0,\, \rho_L^+ \rho_L^-,\, \rho_L^0 \rho_L^0$, 
will do the job from CP-conserving and CP-violating transitions, as we demonstrate below.
The measurement of the time-ordered intensities for these
$16 = 2(f) \times 4(g) \times 2(\text{time ordering})$ combined processes 
is rich in physics and consistencies, 
leading to the relative phase $\gamma$ responsible of direct CP violation. 
The Belle-II experiment at the upgraded KEK facility 
would have the opportunity to perform the analysis presented here 
if enough integrated luminosity is accumulated in the coming years.

The choice of the $\rho_L^{} \rho_L^{}$ channels together with $\pi \pi$ channels
is motivated by their common CP properties, 
as seen in the change of basis~\cite{martin1970elementary} 
from the two-particle states with definite helicity to L-S coupling
\begin{equation}
	\braket{JM;LS}{JM;\lambda_1\lambda_2}
	= \sqrt{\frac{2L+1}{2J+1}}\, C(LSJ|0\lambda) C(s_1 s_2 S | \lambda_1 -\lambda_2)
\end{equation}

\noindent
In particular, for helicities $\lambda_1=\lambda_2=0$, 
the $\rho\rho$ system from $B$ decay is in states with $L=S=0,2$, 
so it has definite symmetry properties under $C=+$, $P=+$ and $CP=+$. 
Therefore, we may use a unified theoretical framework for the discussion of the time-ordered intensities 
associated to the double decays $(f, g)$ and $(g,f)$ 
with decay times $t_1,\, t_2$ such that $\Delta t \equiv t_2 - t_1 >0$. 
For any $g$ decay products, 
the choice of $f=J/\psi K_L$ defines for the $\Delta t$ living partner a CP-forbidden transition, 
whereas $f=J/\psi K_S$ corresponds to a CP-allowed transition. 

Taking the transition amplitude from the $C=-$ entangled $B^0-\bar B^0$ state 
to the time-ordered decay products $f$ and $g$, 
its square and integration over the initial decay time at fixed $\Delta t \equiv t$, 
leads to the double-decay intensity~\cite{Bernabeu:2016sgz}
\begin{equation}
	I(f, g; t) = \frac{e^{-\Gamma |t|}}{16 \Gamma |p q|^2}
	\left|
		e^{i \Delta M t/2} A_H^fA_L^g - e^{-i \Delta M t/2} A_L^fA_H^g
	 \right|^2,
	 \label{eq:Intensity}
\end{equation}

\noindent
with $\Gamma$ the common decay width of the eigenstates 
\mbox{$B_H= p B^0 + q \bar B^0$}, \mbox{$B_L= p B^0 - q \bar B^0$} with definite time evolution, 
$\Delta M = M_H - M_L$ their mass difference and  
$A^{f,g}_{H,L}= \bra{f,g}T\ket{B_{H,L}}$ their decay amplitudes. 
In the absence of CP violation in the mixing for this system, $\abs{p/q} = 1$.
As anticipated,
this intensity presents interference terms between $f$ and $g$,
either direct or through mixing.
With $\Delta \Gamma=0$ for $B_d$ decays, 
there are time-independent and oscillatory terms in $t$ with different physics.
Due to the definite (anti)symmetry of the $C=-$ entangled state, 
Eq.(\ref{eq:Intensity}) satisfies the following expected symmetry property: 
the combined transformation $t \to -t$ and $f \leftrightarrow g$ is the identity. 
Hence the interest in the separate measurements of $I(f,g;t) + I(g,f;t)$ and $I(f,g;t) - I(g,f;t)$
in order to separate even and odd terms in $t$.
As a consequence, 
we find it convenient to express Eq.(\ref{eq:Intensity}) in the basis
$\{ \cos^2(\Delta M\, t/2),\; \sin^2(\Delta M\, t/2),\; \sin(\Delta M\, t) \}$ of time dependencies as
\begin{multline}
	\hat I (f,g;t) \equiv \frac{\Gamma}{\expval{\Gamma_f}\expval{\Gamma_g}}\, e^{\Gamma |t|}\, I(f,g; t) =\\
        \Cd^{fg} \cos^2\frac{\Delta M\, t}{2} + \Cm^{fg} \sin^2\frac{\Delta M \, t}{2} + \Cod^{fg} \sin(\Delta M\, t) \,,
	\label{eq:Ihat}
\end{multline}
where $\expval{\Gamma_f}$ is the average decay probability to $f$ from $B^0$ and $\bar B^0$.
$\hat I(f,g;t)$ is a reduced intensity, with 
$\Cd^{fg} = \Cd^{gf}$,
$\Cm^{fg} = \Cm^{gf}$, and
$\Cod^{fg} = -\Cod^{gf}$ 
the ``intensity parameters'' for each decay pair.
The $\Cd$ parameter shows up since the $t = 0$ separation between the two decays for each $(f,\, g)$ pair,
so it is the signal for a direct correlation between the decay amplitudes.

We introduce the usual Mixing$\times$Decay interference $\lambda \equiv \frac{q}{p}\frac{\bar A}{A}$ from $B^0$ and $\bar B^0$
relevant at any time, 
for each  decay amplitude ---either $\lambda_f$ or $\lambda_g$---. 
In terms of the complex $\lambda$, 
we have the three combinations
\begin{equation}
	C = \frac{1-\abs{\lambda}^2}{1+\abs{\lambda}^2}\,,
	\quad
	R = \frac{2 \Re(\lambda)}{1+\abs{\lambda}^2}\,,
	\quad
	S = \frac{2 \Im(\lambda)}{1+\abs{\lambda}^2}\,,
	\label{eq:CRS}
\end{equation}
with the constraint  $C^2 + R^2 + S^2 = 1$.
The calculation of the intensity parameters in Eq.(\ref{eq:Ihat}) for each double-decay rate  $(f,\, g)$
is then obtained from the combinations (\ref{eq:CRS}) as
\begin{align}
	\nonumber
	\Cd^{fg} &= \frac{1}{2}( 1 - R_f R_g - S_f S_g - C_f C_g ) \, \\
	\label{eq:CdCmCod}
	\Cm^{fg} &= \frac{1}{2}( 1 - R_f R_g + S_f S_g + C_f C_g ) \, \\
	\nonumber
	\Cod^{fg} &= \frac{1}{2}( S_f C_g - C_f S_g ) \,.
\end{align}

Whereas $\Cd$ and $\Cm$ contain real number terms and then select the real part of the time evolution, 
$\Cod$ contains imaginary terms selecting the imaginary part of the time evolution. 
In addition, 
we observe that the time-even parameters are symmetric under the $f \leftrightarrow g$ exchange,
and the odd parameter is antisymmetric, as anticipated.

The decay channels $f = J/\psi K_L, J/\psi K_S$ are known to be well described by their tree level amplitudes, 
which satisfy $\abs{\lambda_f} = 1$, i.e. $C_f = 0$. 
As an important consequence of Eq.(\ref{eq:CdCmCod}), 
a non-vanishing $\Cod$ intensity parameter is trapping penguin amplitudes 
through their modulus contribution $C_g \neq 0$ ($|\lambda_g|\neq 1$) for any decay channel $g$. 
The phase of $\lambda_{f}$ is to a great accuracy the mixing phase $q/p = e^{-2i\phi _{M}}$, 
with $\phi _{M}=\beta$ in the SM. 
Assuming also $\Delta F=\Delta Q$, that is no wrong sign decays, we have
\begin{equation}
	\label{eq:lambdaf}
	\lambda_{S} = -\lambda_{L}=-e^{-2i\phi_{M}}
\end{equation}
imposed by the two opposite CP-eigenvalues for the decay products $J/\psi K_L$ and $J/\psi K_S$. 
The surviving terms in Eq.(\ref{eq:CdCmCod}) are linear in $\lambda_f$, 
implying \textbf{consistency relations} for the absolute and relative normalizations of the intensity parameters,
\begin{equation}
	\Cd^{Lg} + \Cd^{Sg} = 1\,,
	\ \ 
	\Cm^{Lg} + \Cm^{Sg} = 1\,,
	\ \ 
	\Cod^{Lg} + \Cod^{Sg} = 0\,,
	\ \ 
	\forall g,
	\label{eq:normalizationC}
\end{equation}
leading in turn to consistencies for the double-decay time-dependent reduced intensities
\begin{equation}
	\label{eq:normalizationI}
	\hat I(L, g; t) + \hat I(S, g; t) = 1,\;\; \forall g,\; \forall t\,.
\end{equation}
Using the exchange symmetry properties, 
Eqs.~(\ref{eq:normalizationC}, \ref{eq:normalizationI}) are also valid for the time-ordered $(g;\, L,S)$ decays. 
They provide a controlled connection between the CP-forbidden 
and CP-allowed time-dependent transitions for any of the four decay products $g$.

The $\lambda_{g}$ amplitudes $\left( g=\pi \pi ,\rho_{L}\rho_{L}\right) $ can be parameterized as
\begin{equation}
	\label{eq:lambdag}
	\lambda_{g}=\rho_{g}e^{-i2\left( \phi_{M}+\phi_{g}\right) } \,,
\end{equation}
where $\phi_{g}$ is a weak phase in the decay $b\rightarrow u\bar{u}d$, 
and both $\rho_{g}\neq 1$ and $\phi_{g}\neq \gamma $ are due to the penguin contributions. 
At tree level, all $g$ states considered here would have $\phi_{g}=\gamma $.
Notice that the $(R_f R_g + S_f S_g)$ combination appearing in 
\textbf{the $\Cd^{fg}$ intensity parameter is blind to the phase of $q/p$}
and it directly probes
\begin{equation}
	\lambda_{\overset{L}{S}} \lambda_{g}^{\ast} = \pm \rho_{g} e^{i2\phi_{g}}
\end{equation}
where the $\pm$ corresponds to $f = L,\, S$, respectively.
As anticipated, 
no mediation of the mixing is present in the $\Cd$ parameter of the intensity. 
Thus the determination of this direct correlation between the two decay products in Eq.(\ref{eq:Ihat}) 
for these processes becomes
\begin{equation}
	\label{eq:CdLS}
	\Cd^{\overset{L}{S}g} = \frac{1}{2} \left[ 
		1 \mp \frac{2 \rho_{g}}{1+ \rho_{g}^{2}} \cos( 2\phi_{g} ) 
	\right] \,,
\end{equation}
where clearly the mixing is not present.

If penguin contributions were not relevant, 
we would have at tree level
\begin{align}
	\nonumber
	\Cd^{Lg} &= \sin^2 \gamma \; \text{ for all CP-forbidden transitions,} \\
	\Cd^{Sg} &= \cos^2 \gamma \; \text{ for all CP-allowed transitions.}
\end{align}
With the expected $g$-dependent penguin contributions through both $\rho_g$ and $\gamma - \phi_g \equiv \epsilon_g$, 
to be discussed below, 
Eq.(\ref{eq:CdLS}) provides a powerful consistency from the four $g$'s and the two $f$'s 
for the extraction of the CPV $\gamma$-phase. 

Let us focus now on the different information to be accessed 
by the measurement of the other intensity parameters.
In the case of $\Cm$, 
the combination in Eq.(\ref{eq:CdCmCod}) involves the $\lambda_f \lambda_g$ product, 
which connects the $f,g$ decay amplitudes through the mixing.
The use of Eqs.~(\ref{eq:lambdaf}, \ref{eq:lambdag}) leads to
\begin{equation}
	\label{eq:CmLS}
	\Cm^{\overset{L}{S}g} = \frac{1}{2} \left[ 
		1 \mp \frac{2 \rho_{g}^{}}{1+ \rho_{g}^{2}} \cos(4\phi_{M} + 2\phi_{g}) 
	\right] \,.
\end{equation}
The result (\ref{eq:CmLS}) depends on the phase $2\phi_M + \phi_g$, 
indicating explicitly that the $\Cm$ parameter denotes a correlation 
between the two $f$ and $g$ decay channels induced through the mixing. 
As already advertised, 
$\Cm^{L g} + \Cm^{S g} = 1\;\;\forall g$,
as for the other term even in time.

The two $(f,g)$ time-even intensity parameters combine 
in the observable sum of intensities 
for the time-ordered exchange of decay products $f\leftrightarrow g$.
We obtain the result
\begin{align}
	\nonumber
	&\hat I(f,g; t) + \hat I(g,f; t) = \\
	\nonumber
	&= 2\left[ \Cd^{fg} \cos^2(\Delta M\, t/2) + \Cm^{fg} \sin^2(\Delta M\, t/2) \right] =\\
	\nonumber
	&= 1 \mp \frac{2 \rho_{g}}{1 + \rho_{g}^{2}} \left\{ 
		\cos( 2\phi_{g}) \cos^{2}(\Delta Mt/2) 
		\,+ \right.\\
		&\hspace{2.0cm}	\left.
		+\, \cos( 4\phi_{M} + 2\phi_{g} ) \sin^{2}(\Delta Mt/2) 
	\right\} \,,
	\label{eq:Ifg+Igf}
\end{align}
for $f=L,S$ correspondingly. As seen, 
the contributions of the direct CPV phase $\phi_g$ and the mixing-induced CPV phase $2\phi_M + \phi_g$ 
separate in two different time-dependent behaviors, 
the second naturally needing a time slice to become apparent. 
For any of the two $f$ channels and the four $g$ channels, these two terms are separately apparent when
\begin{equation}
 \frac{\Delta M\, t}{2} = n\pi \, ,\ \frac{\Delta M\, t}{2} = (2n+1)\frac{\pi}{2}\,,
\end{equation}
with $n = 0, 1, 2...$\\

The third intensity parameter $\Cod$ can be separated out 
from the difference of the two time-ordered intensities,
\begin{align}
	\nonumber
	&I(f,g;t) - I(g,f;t) = 2\, \Cod^{^L_S g} \, \sin(\Delta M \, t) = \\
	&\hspace{1.5cm}= \mp \left[\frac{1-\rho_g^2}{1+\rho_g^2}\, \sin(2\phi_M) \right] \sin(\Delta Mt) \,,
	\label{eq:Ifg-Igf}
\end{align}
where $\Cod^{Lg} + \Cod^{Sg} = 0 \;\;\forall g$ in this case.
It is worth remarking that 
this intensity parameter would vanish iff the penguin contribution were absent in the $g$ decay channels. 
As the CPV mixing $\sin 2\phi_M$ ($\sin 2\beta$ in the SM) is the best measured parameter in this field, 
Eq.(\ref{eq:Ifg-Igf}) can be used to measure 
the deviation of $\rho_g$ in each of the four $g$-channels from 1, 
induced by the penguin amplitude, 
and check its prediction from the isospin analysis given below. 
Consistently, 
the measurement of observable~(\ref{eq:Ifg-Igf}) for both $f=L$ and $f=S$ has to reproduce a change of sign,
providing in particular the relative normalization of events in these two decay channels.


Besides the factor depending on $\rho_g$, the observables are also affected by the penguin amplitudes in a departure of the phase $\phi_g$ from a common $\gamma /\phi_3$ through
\begin{equation}
	\epsilon_g = \gamma - \phi_g \,,
\end{equation}
to be extracted from a dedicated isospin analysis.
The procedure follows the original ideas of Gronau and London along the path described in Refs.~\cite{Botella:2005ks,Baek:2005cg}. 
The neutral and charged B-meson decays differ in the presence versus absence, respectively, of the penguin contribution to the amplitudes for each final $h = \pi, \rho_L$ system. The charged decay amplitudes $A^{+0} = A(B^+ \to h^+ h^0)$ and $\bar A^{+0} = A(B^- \to h^- h^0)$ have a final $(h^\pm h^0)$ isospin 2 state and, therefore, only the $\Delta I = 3/2$ tree-level amplitude contributes with the weak phase $\gamma$. It is convenient to define, with the same notation for both neutral decay channels $\pi\pi$ and $\rho_L\rho_L$ and using $g = \pm$ or $00$ for the corresponding decay charges,
\begin{equation}
	a_g = \frac{A_g}{A_{+0}} 
	\quad ; \quad
	\overline{a}_g = \frac{\overline{A}_g}{\overline{A}_{+0}},
\end{equation}
in such a way that the double ratio gives
\begin{equation}
	\rho_g \, e^{2i\epsilon_g} = \frac{\overline{a}_g}{a_g} \,.
\end{equation}
The isospin triangular relations with these complex ratios are
\begin{equation}\label{eq:isospin:triangles}
	\frac{1}{\sqrt{2}} a_{+-} = 1 - a_{00} 
	\quad ; \quad
	\frac{1}{\sqrt{2}} \overline{a}_{+-} = 1 - \overline{a}_{00} \,.
\end{equation}
Eqs. \eqref{eq:isospin:triangles} allow to get $a_g$ and $\overline{a}_g$ by using all the branching ratios of the processes $B^\pm \to h^\pm h^0$; $B^0$, $\bar B^0 \to h^+ h^-$, $h^0 h^0$. In Table \ref{tab:isospin_data} we give the summary of our isospin analysis with the present PDG data ~\cite{Zyla:2020zbs}. Taking into account that the $\rho^+_L\rho^-_L$ channel is the one with larger branching ratio, we must conclude that the error in $\epsilon_{\rho^+_L \rho^-_L}$, 
$\delta \epsilon_{\rho^+_L \rho^-_L} = 0.091 = 5.2^\circ $, 
gives us an estimate of the uncertainty due to the present knowledge of the penguin pollution in the determination of $\gamma /\phi_3$. An important improvement in the branching ratios entering in the isopin analysis is expected as an outcome of Belle-II and LHC experiments that will reduce this error significantly.
\setlength{\tabcolsep}{8pt}
\begin{center}
\begin{table}[t]
	\caption{Summary of isospin analyses results.}
	\label{tab:isospin_data}
		\begin{tabular}{ccc}
			\toprule
			$g$		&$\rho_g$	&$\;\;\; \epsilon_g$\\
			\midrule
			$\rho^+_L \rho^-_L$ 	& $1.007\pm 0.076$	& $\;\;\; 0.008\pm 0.091$ \\
			$\rho^0_L \rho^0_L$ 	& $0.972\pm 0.241$	& $\;\;\; 0.007\pm 0.345$ \\
			$\pi^+ \pi^- $ 		& $1.392\pm 0.062$ 	& $\pm(0.307\pm 0.170)$ \\
			$\pi^0 \pi^0 $ 		& $1.306\pm 0.206$	& $\pm(0.427\pm 0.172)$\\
			\bottomrule
		\end{tabular}
		\vspace{0.5cm}
%
	\caption{Benchmark cases used in the numerical simulations.}
	\label{tab:benchmark}
		\begin{tabular}{cccc}
			\toprule
			Benchmark	&$\rho_g$	&$\epsilon_g$	&$g$\\
			\midrule
			$B_{\rho\rho}$	&$1$ 		&$0$		&$\rho^+_L \rho^-_L ,\, \rho^0_L \rho^0_L$  \\
			$B^+_{\pi\pi}$	&$1.35$ 	&$+0.35$	&$\pi^+ \pi^-   ,\, \pi^0 \pi^0$  \\
			$B^-_{\pi\pi}$	&$1.35$ 	&$-0.35$	&$\pi^+ \pi^-   ,\, \pi^0 \pi^0$  \\
			\bottomrule
		\end{tabular}
\end{table}
\end{center}

The intrinsic accuracy of the method proposed in this paper is controlled by our ability to extract $\phi_g$. In order to estimate the expected uncertainty in that extraction, we proceed as follows (further details are provided in the Supplementary Material). First, we fix input values of $\phi_M$ and $\gamma$. For each decay channel $g=\rho^+_L\rho^-_L$, $\rho^0_L\rho^0_L$, $\pi^+\pi^-$, $\pi^0\pi^0$, we also fix input values of $\rho_g$ and $\epsilon_g$, which fix $\phi_g=\gamma-\epsilon_g$, following  the three different benchmark cases in Table \ref{tab:benchmark}. Next, considering the decay channels $f=L,S$, we compute the six coefficients $\mathcal I^{Sg}_{\rm d,m,od}$, $\mathcal I^{Lg}_{\rm d,m,od}$, which control the four time-dependent decay channels $(f,g)$, $(g,f)$, for each $g$. Then, for each $g$, a given number of events is generated according to the four double-decay intensities. The procedure is repeated in order to produce our simulated data, from which $\mathcal I^{Sg}_{\rm d,m,od}$ are extracted including uncertainties, $\mathcal I^{Lg}_{\rm d,m,od}$ are given by eq.\eqref{eq:normalizationC}. Finally $\rho_g$, $\phi_g$, $\phi_M$ are obtained with a simple fit. Notice that the intensity parameters $\mathcal I_{\rm d,m,od}$ depend, respectively, on $\phi_g$, $\phi_M + \phi_g$ and $\phi_M$ phases. Therefore, the inclusion of the $\mathcal I_{\rm m}$ term together with $\mathcal I_{\rm d}$ in the fit allows to avoid the discrete degeneracy $\phi_g \to \phi_g + \pi$, with the information of the quadrant for $\phi_M$. 

We show the results of our analysis in two scenarios A and B taking into account the Belle-II projected luminosity \cite{Belle-II:2010dht,Belle-II:2018jsg,Forti:2022mti} and the corresponding branching ratios: scenario A assumes 1000 $\rho^+_L\rho^-_L$ events of type $B_{\rho\rho}$, 50 $\rho^0_L\rho^0_L$ events of type $B_{\rho\rho}$, 200 $\pi^+\pi^-$ events of type $B_{\pi\pi}^-$ and 50 $\pi^0\pi^0$ events of type $B_{\pi\pi}^-$. Scenario B assumes 500 $\rho^+_L\rho^-_L$ events of type $B_{\rho\rho}$ and 100 $\pi^+\pi^-$ events of type $B_{\pi\pi}^+$. The results of the fit to the generated $\mathcal I^{fg}_{\rm d,m,od}$ in both scenarios are given in Table \ref{tab:Scenarios}.

\setlength{\tabcolsep}{4pt}
\begin{table}[t]
	\begin{center}
		\caption{Results of the fit}
		\label{tab:Scenarios}
		\begin{tabular}{ccccc}
			&\multicolumn{2}{c}{SCENARIO A} 	&\multicolumn{2}{c}{SCENARIO B} \\
			\toprule
			$g$					& $\phi_g$			& $\rho_g$ 		& $\phi_g$			& $\rho_g$ \\
			\midrule
			$\rho^+_L \rho^-_L$ 	& $1.222(020)$	& $1.00(06)$	& $1.222(31)$	& $1.00(08)$ \\
			$\rho^0_L \rho^0_L$ 	& $1.22(09)$ 	& $1.00(24)$ \\
			$\pi^+ \pi^-$ 		& $1.57(12)$ 	& $1.35(12)$	& $0.87(07)$ 	& $1.36(35)$ \\
			$\pi^0 \pi^0$ 		& $1.57(18)$ 	& $1.35(24)$ \\
			\midrule
			&\multicolumn{2}{c}{$\phi_M = 0.384(31)$}	&\multicolumn{2}{c}{$\phi_M = 0.384(40)$} \\
			\bottomrule
		\end{tabular}
	\end{center}
\end{table}
From the results in Scenario A we conclude that,
since $\gamma = \phi_g + \epsilon_g$, 
the error 
$\delta \phi_{\rho^+_L \rho^-_L} = 0.020 =1.1^\circ$ 
gives an idea of the intrinsic statistical limiting error we would expect in the determination of $\gamma$ 
for the assumed number of events.
Combining $\phi _{\rho^+_L \rho^-_L}$ with 
$\epsilon_{\rho^+_L \rho^-_L} = 0.008\pm 0.091$ 
would bring the error in $\gamma$ to the present error in $\epsilon_{\rho^+_L \rho^-_L}$,
hence the importance of its improvement, as already mentioned.
Even before these improvements we can do better 
and fit the three $\mathcal{I}_\mathrm{d,m,od}^{fg}$ for all channels in terms of $\gamma$, $\epsilon_g$ and $\rho_g$ including all the information of the isospin analysis.
In this case the result 
is $\gamma = 1.222\pm 0.080 = \left( 70.0\pm 4.6\right)^\circ$. 
Note that the error on $\gamma$ is smaller than the error in $\epsilon_{\rho^+_L \rho^-_L}$ due to a unique $\gamma$ in all channels,
which presents a quantitative conclusion: 
the present proposal could provide a measurement of $\gamma$ below the $1^\circ$ error 
if the errors in the isospin analysis can be reduced to the level of $\delta\phi_{\rho^+_L \rho^-_L}\simeq 1^\circ$.

For the more conservative scenario B, we get
%
an intrinsic error $\delta \phi_{\rho^+_L \rho^-_L} = 1.8^\circ$. 
Again, using all the information used in the isospin analysis and $\phi_M$, 
we estimate $\gamma = 1.221 \pm 0.085 = \left( 70.0 \pm 4.9 \right)^\circ$,
which reinforces the idea that, with this method, it could be statistically possible to go below $1^\circ$ 
of precision in the determination of $\gamma$, 
thanks to the expected improvements in the data entering in the isospin analysis.

To conclude, with $B^0$--$\bar B^0$ Entanglement, we consider the double decay rate Intensity to flavor-non-specific channels governed by the $c$- and $u$-quarks. It offers a conceptual alternative to the decay of single $B^\pm$ mesons for the extraction of the direct CPV $\gamma/\phi_3$ phase. The needed interference between two decay amplitudes is provided by the exchanged terms of the entangled state and no strong phases appear as essential
ingredients. The 8 time-symmetric $(f, g)$ Intensities with $f =J/\Psi K_L,J/\Psi K_S$, $g = (\pi\pi)^0,(\rho_L\rho_L)^0$ have a tree-level common $\gamma$ phase, $g = \rho^+_L\rho^-_L$ being the benchmark channel. Several constraining consistencies among the different intensities appear. 
We find that an intrinsic accuracy of the order of 1 degree could be achievable for the relative phase of the $f$- and $g$-amplitudes. The present limitation of $\pm 5^\circ$, to be improved by the existing experimental facilities, comes from the phase of the penguin contribution in the $g$-amplitude, extracted from an isospin analysis to neutral and charged B decays.

\begin{acknowledgments}
We would like to thank Francesco Forti and Carlos Mariñas for several discussions about Belle II physics. 
This research has been supported by \emph{Agencia Estatal de Investigación del Ministerio de Ciencia e Innovación} (AEI-MICINN, Spain) Projects PID2019-106448GB-C33 and PID2020-113334GB-I00, \emph{Generalitat Valenciana} Projects GV PROMETEO 2017-033 and GV PROMETEO 2019-113, and the Alexander von Humboldt Foundation. 
MN is supported by the \emph{GenT Plan} from \emph{Generalitat Valenciana}, Project CIDEGENT/2019/024.
\end{acknowledgments}


%

\clearpage
\section*{Supplementary material}
\begin{figure}[ht]
	\centering
	\includegraphics[width=0.475\columnwidth]{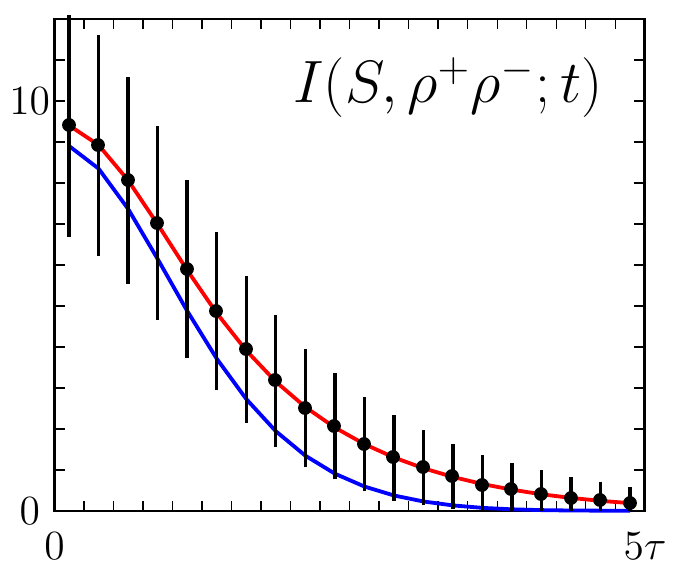}\ 
    \includegraphics[width=0.475\columnwidth]{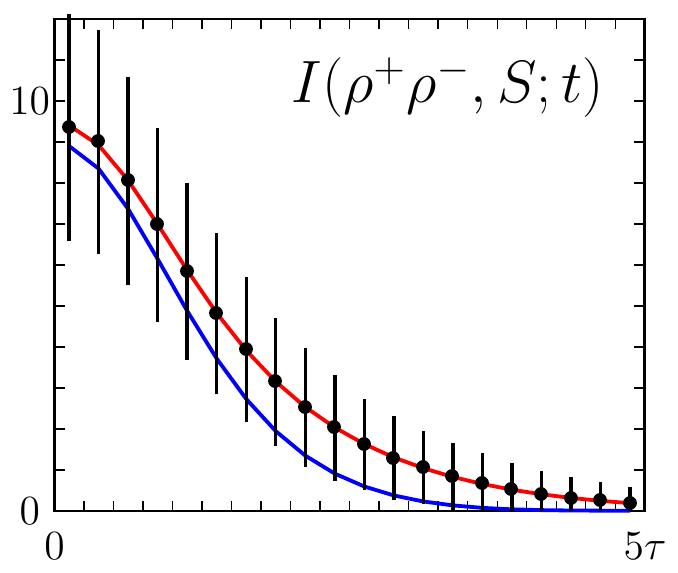}\\
    \includegraphics[width=0.475\columnwidth]{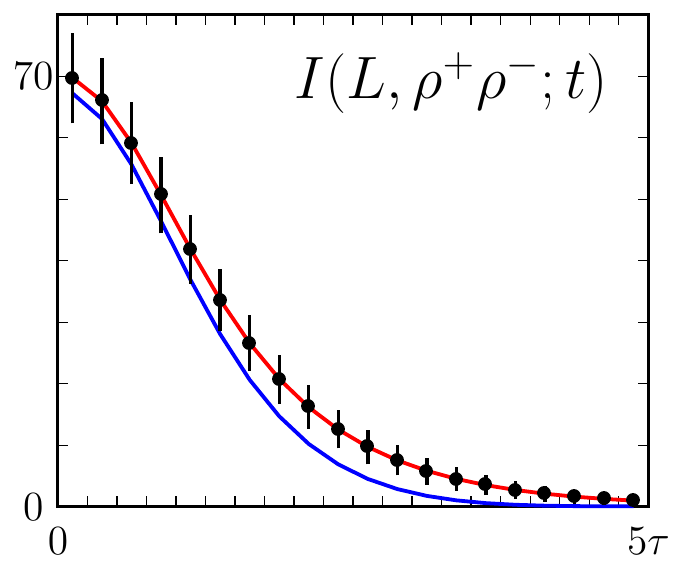}\ 
    \includegraphics[width=0.475\columnwidth]{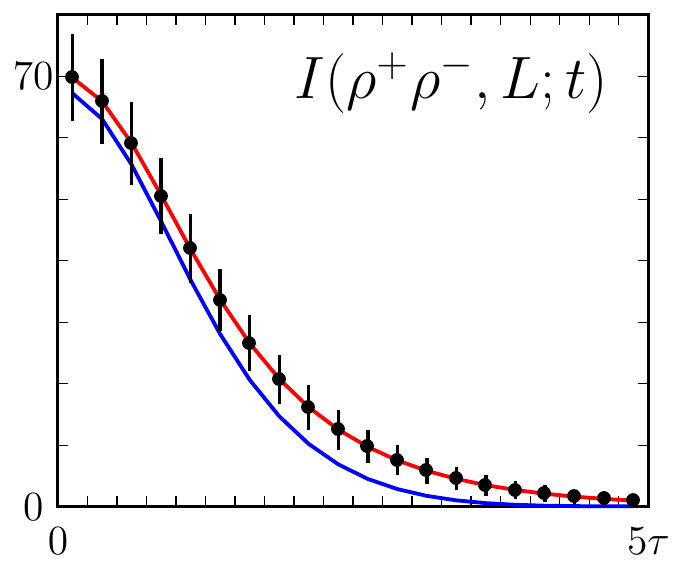}
    \caption{Simulated data, 1000 events, benchmark $B_{\rho\rho}$. Black dots with bars indicate mean values and associated uncertainties; the red curves are the extracted double-decay intensities, while the blue curves correspond to the $\mathcal I^{fg}_{\rm d}$ term in each intensity.} 
	\label{fig:t_bins}
\end{figure}

As discussed in the main text, the intrinsic limitation of the method is controlled by our ability to extract $\phi_g$. The procedure to estimate the expected uncertainty in that extraction is the following.
\begin{enumerate}
\item We fix input values of $\phi_M=\beta=0.384$, $\gamma=1.222$, and, for each decay channel $g=\rho^+_L\rho^-_L$, $\rho^0_L\rho^0_L$, $\pi^+\pi^-$, $\pi^0\pi^0$, we also fix input values of $\rho_g$ and $\epsilon_g$, which fix $\phi_g=\gamma-\epsilon_g$. We consider the three different benchmark cases in Table \ref{tab:benchmark}. \item Considering the decay channels $f=L,S$, the six coefficients $\mathcal I^{Sg}_{\rm d,m,od}$, $\mathcal I^{Lg}_{\rm d,m,od}$ are computed: they control the four time-dependent combinations $(f,g)$, $(g,f)$, for each $g$. 
\item For each $g$, we generate values of $t$, the events, distributed according to the four double-decay intensities. In order to incorporate the effect of experimental time resolution, each $t$ is randomly displaced following a normal distribution with zero mean and $\sigma=1$ ps. Additional experimental effects such as efficiencies are not included. Generation proceeds until a chosen number of events $N_g$ with $|t|\leq 5\,\tau_{B^0}$ has been obtained with the four $(f,g)$, $(g,f)$ combinations altogether. These $N_g$ events are binned. 
\item The procedure is repeated in order to obtain mean values and standard deviations in each bin: these constitute our simulated data, as illustrated in Figure \ref{fig:t_bins}, which corresponds to $g=\rho^+_L\rho^-_L$ (benchmark $B_{\rho\rho}$ in Table \ref{tab:benchmark}), $N_g=1000$ events and 20 bins in $[0;5\,\tau_{B^0}]$. The black dots with bars are the mean values and uncertainties, the red curves are the extracted double-decay intensities, and the blue curves correspond to the $\mathcal I^{fg}_{\rm d}$ term in each intensity. There are no significant differences if one considers, for example, 15 or 10 bins.
\item From the simulated data, one can obtain $\mathcal I^{S\rho^+_L\rho^-_L}_{\rm d}=0.1170\pm 0.0138$, $\mathcal I^{S\rho^+_L\rho^-_L}_{\rm m}=0.1658\pm 0.0456$, and $\mathcal I^{S\rho^+_L\rho^-_L}_{\rm od}=0.000\pm 0.0198$, with $\mathcal I^{L\rho^+_L\rho^-_L}_{\rm d,m,od}$ given by eq.\eqref{eq:normalizationC}, and similarly for decay channels $\rho^0_L\rho^0_L$, $\pi^+\pi^-$, $\pi^0\pi^0$ according to the different benchmarks $B_{\rho\rho}$, $B_{\pi\pi}^\pm$ in Table \ref{tab:benchmark}.
\item Finally we extract $\rho_g$, $\phi_g$, $\phi_M$, with a simple fit to the $\mathcal I^{Sg}_{\rm d,m,od}$.
\end{enumerate}
Concerning the number of events, with the Belle-II design luminosity \cite{Belle-II:2010dht} and the branching ratios $\text{BR}(g)$, $\text{BR}(f)$, we assume that it would be possible to collect 1000 events for $g=\rho^+_L\rho^-_L$, 200 events for $g=\pi^+\pi^-$ and 50 events for both $g=\rho^0_L\rho^0_L$ and $g=\pi^0\pi^0$ channels. We show the results of our analyses for two scenarios.
\begin{itemize}
\item Scenario A assumes 1000 $\rho^+_L\rho^-_L$ events of type $B_{\rho\rho}$, 50 $\rho^0_L\rho^0_L$ events of type $B_{\rho\rho}$, 200 $\pi^+\pi^-$ events of type $B_{\pi\pi}^-$ and 50 $\pi^0\pi^0$ events of type $B_{\pi\pi}^-$. 
\item In scenario B we assume to have 500 $\rho^+_L\rho^-_L$ events of type $B_{\rho\rho}$ and 100 $\pi^+\pi^-$ events of type $B_{\pi\pi}^+$. 
\end{itemize}

\end{document}